SOLAR SYSTEM

# Sifting through the debris

Asantha Cooray

*A quadrillion small bodies beyond Neptune, spotted through the occasional X-ray dimming of a distant source, had previously gone unnoticed. Models of the dynamics of debris in the Solar System's suburbs must now be reworked.*

For a long time, the Solar System was believed to consist of the Sun, the Moon and the six inner planets — including Earth — as far as Saturn. In the past few centuries since the invention of the telescope, we have added other gas giant planets (Neptune and Uranus), moons for the other planets, the asteroid belt and, in 1930, Pluto. On page XXX of this issue, Chang and colleagues[1] present observational estimates of the make-up of the Solar System's most distant components — the 'trans-neptunian' objects — that require a rethink of models of how these bodies formed.

It was in 1943 that Kenneth Edgeworth suggested that the outer Solar System might contain a large number of planetesimal objects[2]. These are the residual debris from the planetary formation process that was swept off to the edge of the Solar System by the gravity of the giant planets. Since 1992, imaging searches for slow-moving faint objects have revealed more than 50,000 planetoids, which are generally hundreds of kilometres across, orbiting beyond Neptune[1]. But despite theoretical models that had suggested their existence, there had been no conclusive evidence for their presence of smaller bodies.

Chang *et al.*[1] change that. Their observations indicate that there are in fact far more small planetesimals from ten to a hundred metres across than theorists had predicted in the trans-neptunian debris field. The authors infer this from random, rapid dips in the intensity of Scorpius X-1, the brightest X-ray light source in the sky that was monitored by NASA's Rossi X-ray Timing Explorer over a long period. The passage of a dark foreground object within the Solar System across the background source viewed from Earth is thought to cause such a dip in brightness in the process known as occultation. An object 100 metres across at the edge of the Solar System could cause an occultation of a few milliseconds' duration. Such a small object so far away cannot be directly imaged with even the largest telescopes, as the amount of sunlight that it reflects back to Earth is miniscule.

The use of stellar occulations to reveal unknown facets of our Solar System is not new. Uranus's rings were discovered in this way, as were a host of details about the atmospheres of other planets and their satellites[4]. The idea of using occultations to search for small bodies in the outer Solar System has been around since 1976 (ref. 5). But in practice, most detectors used for astronomical observations at optical wavelengths cannot record changes at the sub-second time intervals required for this particular hunt.

Chang and colleagues' inspiration was to realize that detectors counting X-ray photons record light at shorter, millisecond timescales. So instead of looking at distant optical light coming from stars, they used an existing long-duration observation of Scorpius X-1, which lies close to the ecliptic plane. Based on the number of rapid dips in brightness on timescales of a few to 10 milliseconds, and assuming that the occulting bodies are about as distant as Pluto, they estimate that the total number of bodies between 10 and 100 metres across in the outer Solar System is more than a quadrillion, or $10^{15}$ (see Box 1).

That number suggests that there is a 20-metre-sized planetesimal in every volume of radius 100,000 kilometres in the outer Solar System. Astronomically speaking, that is extreme overpopulation. The debris field resembles a crowded dance floor where, inevitably, frequent collisions occur. These collisions break up the bodies, grinding them down to smaller objects that are eventually removed from the Solar System. Computer models indicate that there should currently be between a billion ($10^9$) and a trillion ($10^{12}$) bodies between 10 and 100 meters in size[6], far less than Chang and colleagues' number[1]. This discrepancy implies that collisions are less frequent than thought. As faster moving objects collide more frequently, these small bodies are likely to be moving slowly relative to the larger planetesimals in the disk.

The occultation technique has an inherent problem: a dip does not establish the exact geometry of an event, as a small body nearby will produce a similar dip as a large object further away would. Fortunately, one can establish the distance to an occulting body independently of its size. One does this by observing the tiny modifications in a dip's shape as the light beam from the background source diffracts over the surface of the foreground body[7]. But despite the high X-ray luminosity of the Scorpius X-1 source, it was not bright enough to allow the study of these expected small changes. A dedicated space-based monitoring mission, optimized for precision stellar occultation measurements at millisecond time intervals, could search for diffraction patterns to establish the exact extent of the debris field[8]. Whipple, a Discovery mission recently proposed and now under consideration for funding by NASA, can monitor for occultations of background optical stars by foreground small bodies as far as the hypothesized Oort cloud at the extreme edge of the Solar System.

The Taiwan–America Occultation Survey[9] currently operates four robotic optical telescopes in central Taiwan that monitor optical light from stars at 0.2-second intervals. It will soon make a census of the number of kilometre-sized bodies on the outer edges of the Solar System. The cumulative thermal 'blackbody' radiation emission from planetesimals could also soon be revealed in exquisite far-infrared images from the European Space Agency's Herschel and Planck missions, which are now scheduled for launch in 2008. Meanwhile, theorists must scrutinize their models to explain the dense debris field at the fringes. Why it is not sparse may hide important clues to the mass and extent of the primordial gas disk from which the Solar System formed.

*Asantha Cooray is in the Department of Physics and Astronomy, University of California, Irvine, California 92697, USA.*

## Box 1: Big ones, small ones, ones of every size

Before Chang and colleagues' results[1], we knew only of the existence of planetoids with sizes greater than a few tens of kilometres in the outer Solar System. These objects were found by largest telescopes in the world looking for faint sources moving slowly relative to the fixed stars. The number of bodies with sizes greater than about 100 kilometres is now estimated to be about 50,000 (regions marked in green and blue).

Chang and colleagues' X-ray occultation technique allows them, for the first time, to see smaller objects. They estimate the total number of small objects between 10 and 100 metres in size to be around $10^{15}$ (orange region). That is roughly 1,000 to 100,000 times more than computer models indicate for those sizes (black and red dotted lines)[6].

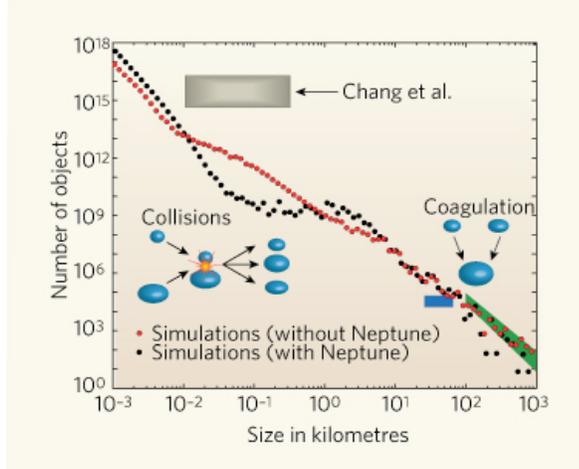

These computer calculations involve two physical processes. At early times, small bodies coagulate to form larger ones; and as the system evolves, fast moving, smaller bodies collide with slower, larger ones. Neptune, acting as a gravitational stirrer, interacts with these small bodies and is primarily responsible for increasing their speed during Neptune's migration outward early in the solar system history.

The system of 'trans-neptunian objects' is still evolving today, with collisions slowly grinding small objects to dust. To explain the discrepancy between the computer models and Chang and colleagues' result, the small bodies might need to move slower than the larger bodies to avoid frequent collisions. Another possible resolution is to extend the region of small bodies to a significantly further beyond Pluto, as the observed occultation dips are not precise enough to establish the distances to each planetesimal independently of its size.